# Face Coverings, Aerosol Dispersion and Mitigation of Virus Transmission Risk

**Ignazio Maria Viola,**[1] **Brian Peterson,**[1] **Gabriele Pisetta,**[1] **Geethanjali Pavar,**[1] **Hibbah Akhtar,**[1] **Filippo Menoloascina,**[1] **Enzo Mangano,**[1] **Katherine E. Dunn,**[1] **Roman Gabl,**[1] **Alex Nila,**[2] **Emanuela Molinari,**[3] **Cathal Cummins,**[4] **Gerard Thompson,**[5] **Tsz-Yan Milly Lo,**[6,7] **Fiona C. Denison,**[8] **Paul Digard,**[9] **Omair Malik,**[10] **Mark J. G. Dunn,**[11] **Catherine M. McDougall,**[6] **Felicity V. Mehendale**[7]

[1] School of Engineering, University of Edinburgh
[2] Lavision UK Ltd.
[3] School of Informatics, University of Edinburgh
[4] Maxwell Inst. for Math. Sciences, Dep. of Math., and Inst. for Infrastructure & Env., Heriot-Watt University
[5] Centre for Clinical Brain Sciences, Edinburgh
[6] Paediatric Critical Care Unit, Royal Hospital for Sick Children, Edinburgh
[7] Usher Institute, University of Edinburgh
[8] The Queen's Medical Research Institute, Edinburgh
[9] The Roslin Institute, University of Edinburgh
[10] Department of Anaesthesia, Royal Hospital for Sick Children, Edinburgh
[11] Department of Critical Care, NHS Lothian

CORRESPONDING AUTHOR: Ignazio Maria Viola (e-mail: i.m.viola@ed.ac.uk)

This work received funds from the European Research Council (grant no. 759546), the Engineering and Physical Sciences Research Council (EP/P020593/1, EP/L016680/1) and the Institute Strategic Programme Grant support (no.BB/P013740/1) from the Biotechnology and Biological Sciences Research Council UK.
This article has supplementary downloadable material

**ABSTRACT** The SARS-CoV-2 virus is primarily transmitted through virus-laden fluid particles ejected from the mouth of infected people. Face covers can mitigate the risk of virus transmission but their outward effectiveness is not fully ascertained. Objective: by using a background oriented schlieren technique, we aim to investigate the air flow ejected by a person while quietly and heavily breathing, while coughing, and with different face covers. Results: we found that all face covers without an outlet valve reduce the front flow through by at least 63% and perhaps as high as 86% if the unfiltered cough jet distance was resolved to the anticipated maximum distance of 2-3 m. However, surgical and handmade masks, and face shields, generate significant leakage jets that may present major hazards. Conclusions: the effectiveness of the masks should mostly be considered based on the generation of secondary jets rather than on the ability to mitigate the front throughflow.

**INDEX TERMS** COVID-19 pandemic, face coverings, face masks, aerosol dispersal, aerosol generating procedures.

**IMPACT STATEMENT** These results show the effectiveness of face coverings in mitigating aerosol dispersion and can aid policy makers to make informed decisions and PPE developers to improve their product effectiveness.

## I. INTRODUCTION

It is now ascertained that the use of face coverings is paramount to mitigate SARS-CoV-2 virus transmission and to address the COVID-19 pandemic [1]. Several studies investigating mask efficacy have been undertaken in recent months, using different measurement techniques and numerical models [2]–[11]. Yet, due to the multi-faced nature of this problem, we still do not have a complete understanding of the flow and around different face coverings and their relative effectiveness in mitigating droplets and aerosol dispersal, and virus transmission.

Since the very early stages of the pandemic, several researchers have investigated the filtering efficacy of different fabrics. Rodriguez-Palacios et al. [2] measured the distance travelled by droplets (20 µm - 900 µm) passing through different fabrics. They used spray bottles filled with an aqueous suspension and found that all masks were highly effective in mitigating droplet dispersion and the contaminated area. Wang et al. [3] focused on the pressure difference generated by different fabrics and their bacterial filtration efficacy. They found that three double-layers of any tested fabric could meet the pressure difference standards and particle filtration efficiency of graded masks. More recently, Ueki et al. [4] tested two manikin heads in a biosafety level 3 box, exhaling droplets that were originally produced by human respiration and coughs and that were loaded with infectious SARS-CoV-2. Remarkably, they found that a face mask worn by the source is more effective in mitigating virus transmission than a mask worn by the recipient. Asadi and co-workers [5] confirmed the outward effectiveness of face masks by testing human volunteers. They found that the outward emission of micron-scale aerosol particles is reduced by 90% and 74% for speaking and coughing, respectively, when the source wears a mask. These results were complemented by Bandiera et al. [6] who focused on the largest (10 μm -1000 μm) droplet. They tested both a speaking and coughing simulator as well as human volunteers, and found that even a single layer cotton mask blocks 99.9% of large droplets.

In parallel to studies on the filtering effectiveness of masks and fabrics, research has been undertaken to characterise the flow field and the fluid mechanics of wearing a mask. Pioneering studies on this subject were presented in a series of papers co-authored by Settles and Tang [12]–[16], who used a schlieren optical method to visualise the airflow emitted by coughing. Tang et al. [16] showed the airflow ejected from a human person coughing, unfiltered as well as when the person is wearing an N95 or a surgical mask. They concluded that the N95 masks block the formation of the jet while the surgical mask redirects it sideways. Qualitative visualisations of the airflow around a person wearing a face cover were recently performed by Verma et al. [7], [8]. They used a laser sheet to illuminate particles in suspension, resulting in impressive visualisations rich in details of the flow structures. Their results highlight the need for further investigate leakage jets from loosely fitted masks and warn on the potentially long distances travelled by unfiltered coughs (>6 ft) and air jets from valves and around shields.

The above studies did not aim to provide quantitative measurements of the flow field nor of mask effectiveness. In contrast, the first quantitative analysis of the flow field was provided by Kähler and Hain [10], who performed particle image velocimetry (PIV) of the air jet exhaled by human volunteers with and without a face mask. The authors concluded that all face masks offer a good outward protection because of the flow resistance that slows down the exhaled air jet. Additional analysis by the authors was dedicated to inward protection for the wearer. Dbouk and Drikakis [11] provided complementary quantitative information on the flow field around face masks using computational fluid dynamics simulations, which informed on the air jets and the droplet concentration on the complete tridimensional volume around the source. As expected, this study highlighted the extent and the importance of the leakage jets previously observed experimentally by, for example, Tang et al. [16] and Kähler and Hain [10].

The present paper complements the work of Kähler and Hain [10] and Dbouk and Drikakis [11] using an experimental technique similar to the pioneering work of Tang et al. [16].

We use background oriented schlieren (BOS), also known as synthetic schlieren [17], which visualises density air gradients. With this technique, we quantify the velocity and direction of the exhaled jets for a wide range of face coverings: FFP1, FFP2, respirator, surgical mask, handmade mask, lightweight 3D printed face shield with visor and heavy-duty commercial face shield. We characterise the different face coverings for the different leakage jets for quiet and heavy breathing, and coughing. We found that any of these face coverings decreased the flow through by at least two-thirds. We used both a human volunteer and an anatomical correct manikin connected to a cough simulator. The latter enables highly reproducible tests to compare different face covers. With the aid of two intensive care specialists, we also used the manikin to simulate the extubation of a patient and we visualise the ejected cough, which revealed the virus transmission risk of this common procedure of COVID-19 patients.

## II. RESULTS

BOS images were used to compute the flow velocity, the spread angle of the air jet and its direction. The jet direction is given by the angle that it forms with the horizon, taken positive rotating anticlockwise (Fig. 1). The jet direction is the bisector of the spread angle, which is the angle between the visible boundaries of the jet. Each test was repeated between 3 and 10 times for each test with face coverings and unfiltered, respectively, ensuring that the 95% confidence interval (95% CI) was within $\pm 5°$ and $\pm 2$ cm. The reported distances travelled by the jets are based on the projections of these jets on the focus plane of the camera.

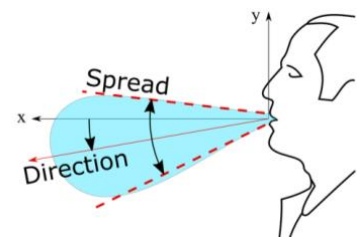

Fig. 1. Reference system and definition of spread and direction of exhaled air.

### A. Unfiltered Quiet and Heavy Breathing and Coughing

With our experimental setup we could observe the thermal plume generated by the person. The closest layer of air to the body is warmer and lighter than the surrounding air and thus it moves upwards as a thermal plume. The expiration begins with a gentle air jet that has a low momentum and it is redirected upwards together with the plume by the buoyancy. Successively, when the ejected flow has a higher momentum, it is displaced in a straight direction. This direction is typically slightly lower than the horizon (Fig. 1) but it depends on the face geometry [18]. Figure 2 shows this first stage of the expiration, where there is both the novel straight jet pointing slightly downwards and the initially exhaled air that has been diverted in a vertical plume and that has been displaced by the front of the new jet. As the expiration phase progresses, the air jet becomes more horizontal and clearly extends beyond the boundary of the field of view at 562 mm from the mouth.

Heavy breathing has a shorter duration and a higher frequency than quiet breathing, with a nine-fold increase in velocity and a three-fold increase in volume flux (Table 1). The higher inertial force results in a straight jet that also clearly extends well beyond the boundary of the field of view, 551 mm from the mouth (Fig. 3).

Table 1. Expiration parameters measured for the quiet and heaving breathing tested conditions.

|  | **Duration** (s) from spirometry | **Average speed** (m/s) from BOS | **Total volume exhaled** (ml/breath) from spirometry |
|---|---|---|---|
| Quiet breath | 3.2 | 0.5±0.2 | 1290 |
| Heavy breath | 2.1 | 5.0±0.8 | 4640 |

The airflow generated by coughing generates an airflow (Fig. 4) that is typically twice as fast as heavy breathing, where the maximum velocity is experienced at the very beginning of the event due to the explosive release from the glottis. During coughing, we observed a puff that is similar but less uniform in speed than the jet generated by breathing heavily, and that is still well discernible 1 m away from the mouth. While the airflow generated by quiet and heavy breathing is a jet with a continuum source of momentum, the puff is a vortex ring [19] whose original source of momentum has ceased while the puff travels forward.

From the observation of the videos (e.g. Supplementary Materials, Test 253), it can be seen that the puff initially travels straight or slightly downwards, driven by inertia (Test 253, Frame 86), and successively the angle reduces (Test 254, Frame 72) when the buoyancy is no longer negligible compared to the inertia. This trend is consistent with observations from other authors [20], [21].

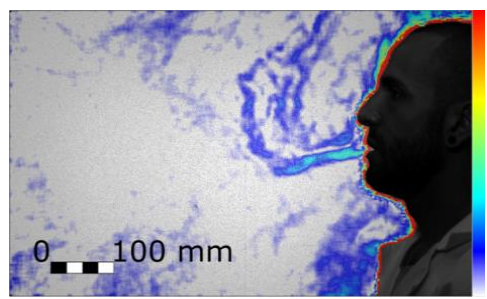

Fig. 2. Initial stage of the quiet breathing expiration (Test 204, Frame 490). Colour bar from blue to red shows low to high density gradient.

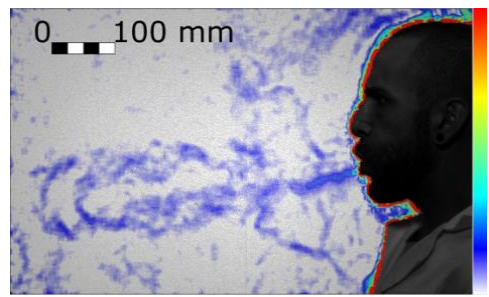

Fig. 3. Fully developed heavy breathing jet (Test 198, Frame 179). Colour bar from blue to red shows low to high density gradient.

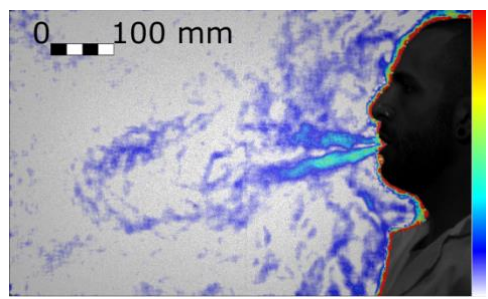

Fig. 4. Cough of a real person (Test 188, Frame 55). Colour bar from blue to red shows low to high density gradient.

### B. Effectiveness of Different Face Covers

Face coverings were tested with the manikin to ensure high repeatability. For the unfiltered tests, the differences between a human cough and that of the manikin are summarised in Table 2 (see also Supplementary Materials, Spirometry). The differences are well within the variability observed between different people [22]–[24]. The different postures are the main reason for the differences in puff direction.

Table 2. Comparison between the cough jets of the real person and the manikin.

|  | **Cough peak speed** (m/s) from BOS | **Jet spread** (deg) | **Direction** (deg) |
|---|---|---|---|
| Real person | 8 | 30±5 | -6±5 (downwards) |
| Manikin | 8 | 30±5 | -17±5 (downwards) |

We found that the exhaled air dispersal for quiet and heavy breathing, as well as for coughing, is significantly mitigated by any face cover. For example, Fig. 5 and 6 show the manikin coughing and wearing an FFP2 and a handmade mask, respectively. For every breathing and coughing condition, the difference between with and without face covering is always significantly larger than the differences between any tested face covering.

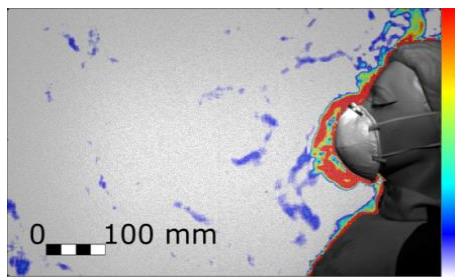

Fig. 5. Cough airflow dispersion contained by an FFP2 mask, which showed the most effective prevention of the frontal throughflow (Test 258, Frame 75). Colour bar from blue to red shows low to high density gradient.

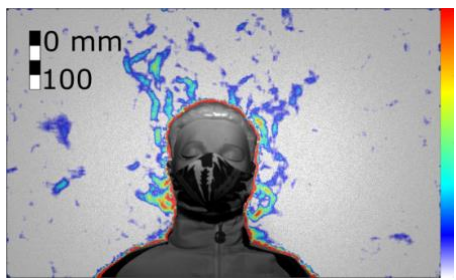

Fig. 6. Cough airflow dispersion partially contained by a handmade mask allowing significant leakage jets (Test 266, Frame 59). Colour bar from blue to red shows low to high density gradient.

Figure 7, 8 and 9 show the front throughflow direction and distance travelled for quiet breathing, heavy breathing and coughing, respectively. The unfiltered quiet breathing and heavy breathing ("no mask" in Fig. 7 and 8), as well as the redirected respirator airflow in the coughing test (Fig. 9), extended beyond the boundaries of the field of view.

We distinguish between unfiltered flow in the absence of face covering (yellow solid lines in Fig. 7-9), filtered flow that flows through the fabric of the mask (red dashed lines), and redirected flow such as that of the respirator and the face shields (blue dotted lines). The redirected flow is partially filtered because the largest droplets, which follow a ballistic trajectory, are likely to land on the surface of the surface of the respirator valve or the face shield.

The data shown graphically in Fig. 7, 8 and 9 is also reported in Supplementary Material, Tables SM-1, SM-2 and SM-3. These tables include also the direction of distance travelled of the largest leakage jet, which is brow-ward for quiet and heavy breathing, and crown-ward for coughing. The definitions of the leakage jets are presented in Fig. 10 and 11 and are discussed in more details in Section IIC.

Between those that we tested, the FFP2 mask was the most effective face covering in mitigating all exhaled air dispersal. A key issue of the FFP2 mask is that it must be shaped to the nose to ensure a proper sealing. When a good sealing is not achievable, we observed a crownward leakage jet displaced beyond the upper boundary of the field of view (Supplementary Material, Table S3). Conversely, when correctly sealed, a leakage was not observed (e.g. Supplementary Materials, Tests 20, 22, 103, 104, 119, 120, 145, 146).

FFP1 was the second most effective face cover in mitigating leakage and the displacement of the front throughflow for quiet and heavy breathing (Fig. 7 and 8), respectively. However, the main weakness of FFP1 is the poor protection while coughing compared to FFP2 (Fig. 9).

On the other hand, the respirator did not mitigate the displacement of the front throughflow (Fig. 7, 8 and 9). In fact, it has a valve system that filters the inhaled air, but it does not filter the exhaled air. The frontal jet is simply redirected downwards, and minimal crown leakage jet is observed in the case of explosive events such as coughs.

The other face covers (surgical and handmade masks, and the shields) showed mixed performances, but the handmade mask was the least effective in stopping air leakage (Supplementary Material, Table S2 and S3). It is important to remark that there is a wide range of handmade masks and thus care must be used in generalising these results. Figure 6 shows a frontal view of the manikin coughing while wearing the handmade mask. While the travelled distance of the front throughflow was effectively reduced as for the other masks ($15\pm2$ cm, Fig. 9), this mask led to multiple leaking jets that could extend upwards, downwards and backwards quiet significantly (see Section IIC).

Face shields typically block the front throughflow, but some airflow was found to leak through seams and joints and be displaced horizontally by few centimetres. This is the case, for example, of the heavy-duty commercial face shield during heavy breathing (Fig. 8). Of course, face shields also generate upwards, downwards, sideway, and particularly strong backwards leakage jets (see Section IIC).

C. *Different Leakage Types*

The different leakage types are shown in Fig. 10 and 11, and their longest travelled distances are reported in Table 3 and 4. A dangerous leakage jet is the backward jet from surgical masks. Air escapes from the side of the mask and it is projected backwards at high speed, potentially resulting in a significant displacement. The backward jet produced by a person breathing with a surgical mask and it extends beyond the end of the field of view at 193 mm from the back of the head (Table 4). This jet is produced by every face cover but not by the FFP1 and FFP2 masks, and by the respirator (Table 4). It is particularly pronounced for surgical masks. Often times, the leakage from the side jet contributes to the backward jet.

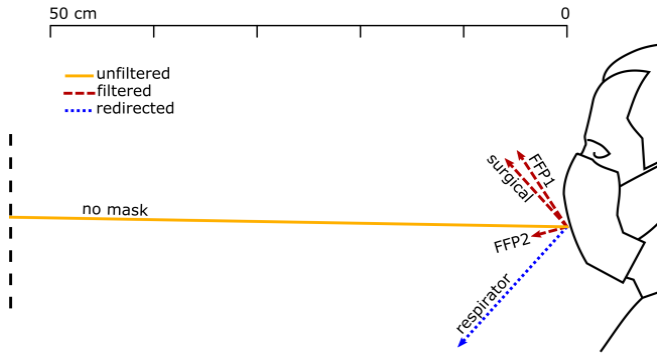

Fig. 7. Direction and distance travelled of the front throughflow for a person quiet breathing with different face coverings. The solid yellow line shows the unfiltered air exhaled by a person not wearing a face cover. Red dashed lines show filtered air that flew through the mask fabric. Blue dotted lines show the only partially filtered air redirected by the respirator valve.

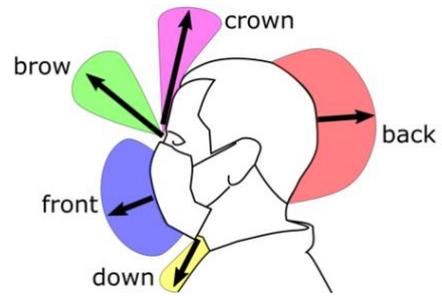

Fig. 10. Front view schematic of the main leakage jets generated by the different face covers.

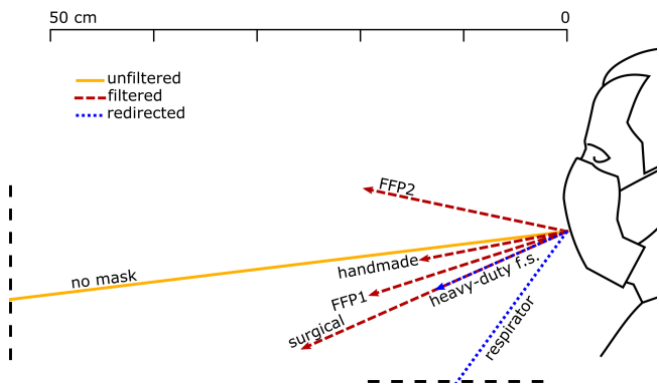

Fig. 8. Direction and distance travelled of the front throughflow for a person heavy breathing with different face coverings. The solid yellow line shows the unfiltered air exhaled by a person not wearing a face cover. Red dashed lines show filtered air that flew through the mask fabric. Blue dotted lines show air redirected by the respirator valve and leaked through seams and joints of the heavy-duty face shield.

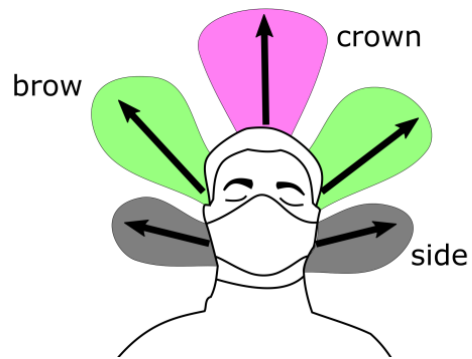

Fig. 11. Side view schematic of the main leakage jets generated by the different face covers

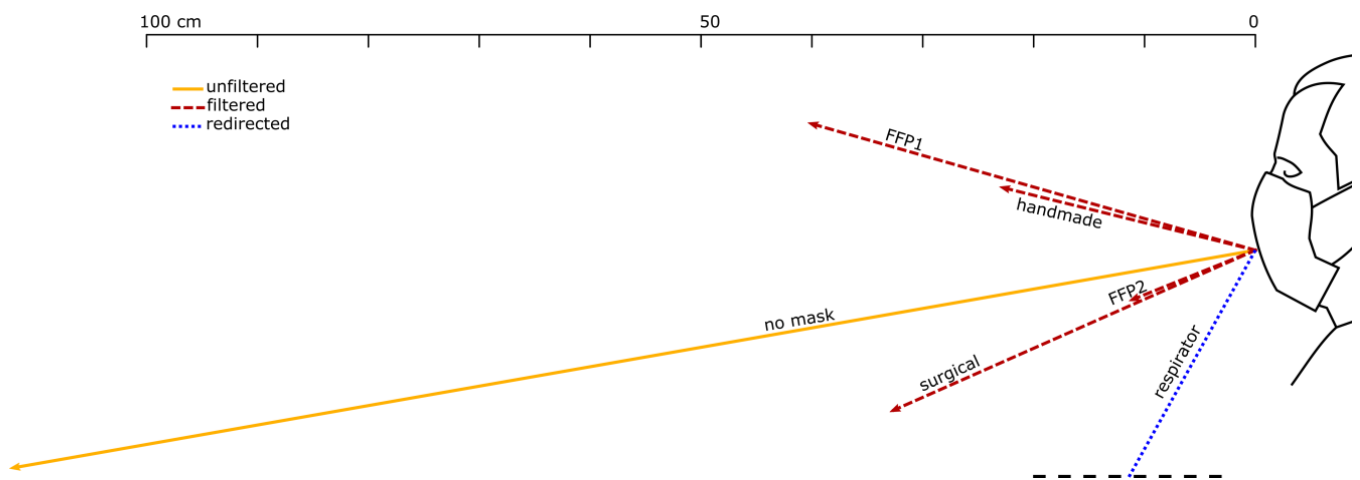

Fig. 9. Direction and distance travelled of the front throughflow for the manikin coughing with different face coverings. Unfiltered air, filtered air that flew through the mask fabric, and air redirected by a valve, are indicated by a solid yellow line, red dashed lines, and a blue dotted line, respectively.

**IEEE Open Journal of Engineering in Medicine and Biology**

*Science*

**Table 3**. Maximum distance travelled by jets for different masks (95% CI < ±2 cm).

| Mask | Front throughflow | | | | Brow | | | | Down | | | |
|---|---|---|---|---|---|---|---|---|---|---|---|---|
| | Flow | Test | Frame | Dist. (cm) | Flow | Test | Frame | Dist. (cm) | Flow | Test | Frame | Dist. (cm) |
| surgical | cough | 256 | 77 | >36 | heavy | 10 | 186 | >27 | not discernible | | | |
| FFP1 | cough | 257 | 236 | 42±2 | heavy | 18 | 56 | 18±2 | not discernible | | | |
| FFP2 | heavy | 104 | 95 | 20±2 | not discernible | | | | not discernible | | | |
| handmade | cough | 259 | 123 | 23±2 | heavy | 106 | 32 | 20±2 | heavy | 29 | 43 | >15 |
| respirator | cough | 260 | 59 | 31±2 | not discernible | | | | not discernible | | | |
| heavy-duty commercial face shield | heavy | 110 | 206 | 14±2 | not discernible | | | | heavy | 88 | 96 | >47 |
| UoE lightweight 3DP face shield+ opaque cover | not discernible | | | | heavy | 132 | 94 | >16 | heavy | 208 | 29 | >30 |

**Table 4**. Maximum distance travelled by jets for different masks (95% CI < ±2 cm).

| Mask | Crown | | | | Side | | | | Back | | | |
|---|---|---|---|---|---|---|---|---|---|---|---|---|
| | Flow | Test | Frame | Dist. (cm) | Flow | Test | Frame | Dist. (cm) | Flow | Test | Frame | Dist. (cm) |
| surgical | heavy | 99 | 45 | >22 | heavy | 116 | 102 | 19±2 | heavy | 99 | 95 | >19 |
| FFP1 | cough | 257 | 64 | >17 | cough | 264 | 146 | 16±2 | not discernible | | | |
| FFP2 | cough | 258 | 74 | >19 | cough | 265 | 225 | 14±2 | not discernible | | | |
| handmade | heavy | 106 | 35 | >22 | cough | 266 | 139 | 18±2 | quiet | 105 | 195 | >18 |
| respirator | cough | 260 | 57 | >11 | not discernible | | | | not discernible | | | |
| heavy-duty commercial face shield | cough | 261 | 70 | >26 | heavy | 128 | 66 | 18±2 | heavy | 110 | 70 | >20 |
| UoE lightweight 3DP face shield + opaque cover | cough | 262 | 80 | >10 | cough | 269 | 92 | 8±2 | heavy | 134 | 90 | >15 |

In the videos available in the Supplementary Materials (e.g. Tests 99, 105, 106, 110, 134), this backward jet appears later, and it diffuses more rapidly than the other jets. We observe two overlaid backward jets originated from the left and the right side of the person. This explains the comparatively high strength and diffusivity of this jet compared to the other jet types. This jet type is more often observed while heavy breathing (Table 4), suggesting that it could be a key hazard from runners.

The brow and crown jets are directed upward through an opening at the top of the mask. These jets are often dispersed within with the thermal plume associated with a natural convection boundary layer travelling vertically upwards from the human body as described by Tang et al. [16]

Another potentially dangerous leakage jet that should be carefully considered is the downward jet. It is generated by all face covers but not by the FFP1 and FFP2 masks, and by the respirator. Figure 12 shows this jet for the lightweight 3D-printed face shield. Different shield shapes would result in different jets but, unless curved below the chin, they are likely to generate an intense and fast downward jet (13 m s$^{-1}$ in Fig. 12). The lightweight 3D-printed face shield is completely open at the bottom permitting free passage of air and it appears (from the signal intensity and initial jet velocity) that most of the ejected air is redirected downwards. The heavy-duty commercial face shield is curved on the four edges, including under the chin, resulting in a less intense downward jet with about half of the initial speed than the lightweight 3D-printed face shield.

### D. Aerosol Generating Procedures

Two intensive care specialists mimicked the intubation and extubation of a patient using the manikin. During the entire procedure, the clinicians are positioned behind the head of the patient, who lies supine on a bed. The manikin was intubated with a Portex size 6.5 endotracheal tube (internal diameter 6.5 mm, external diameter 8.9 mm), secured at a depth of 23 cm from the lips. The part of the tube that remains outside of the mouth is secured in place with tape attached to the patient's face. To prevent airflow bypassing the tube, there is an inflatable balloon cuff inflated at 1.5-2.5 cm from the vocal cords [25], [26] once the tube is in place, sealing the space between the external surface of the tube and the mucosa of the trachea.

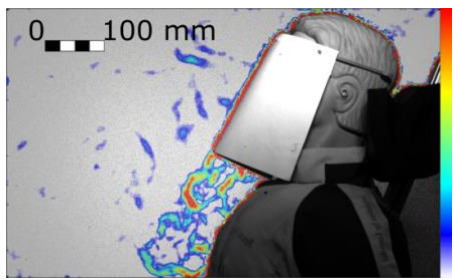

Fig. 12. Cough with the UoE lightweight 3D-printed face shield resulting in a strong downwards jet (Test 262, Frame 77). Colour bar from blue to red shows low to high density gradient.

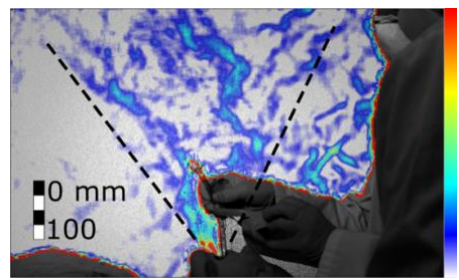

Fig. 14. Cough during the extubation procedure with hands interference (Test 279, frame 538). Colour bar from blue to red shows low to high density gradient.

The extubation procedure takes typically between 10 s and 15 s and was fully recorded. Firstly, the clinicians remove the tape. While a clinician holds the tube steady, the other deflates the sealing cuff. The tube is then extracted slowly. The patient typically coughs repeatedly while the tube is pulled out. Once the tube is removed, a clinician inserts a suction catheter into the patient's pharynx to vacuum liquid residuals and retained secretions to minimise aspiration – the contamination of the lower airways with secretions. The manikin coughs 10 times, each cough lasts 0.2 s spaced by a 0.2 s pause.

We found that when the patient coughs with the tube, the jet of expired air is ejected both from the mouth and from the disconnected tube. Since the tube can have different orientations and might not be aligned with the jet emitted from the mouth, the tube increases the directions toward which the cough is spread. Figure 13 and 14 show the combined jet through the mouth and the tube during the extubation. The jet is not directed straight upwards or towards the clinicians, but at an angle (73°±5°) towards the patient's feet. However, when the clinicians placed their hands through the jet (Fig. 14), the jet spread over a wider angle (from 32°±5° to 46°±5°) in a more upright direction (81°±5°) towards the clinicians.

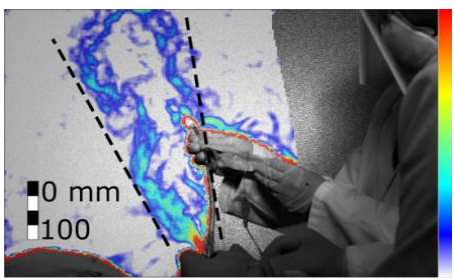

Fig. 13. Cough during the extubation procedure without hands interference (Test 278, frame 738). Colour bar from blue to red shows low to high density gradient.

### III. DISCUSSION

Without face covers, we show that a turbulent air jet extends straight driven by inertia from the mouth of the person in an almost horizontal direction. The jet gradually increases in size and the velocity decreases, and when the inertial force is comparable to the buoyancy force, it bends upwards. Face coverings are found to be highly effective in decreasing the velocity of the jet and, in turn, the horizontal distance travelled by the aerosol. This equally applies to the jets generated by quiet and heavy breathing, and to the puffs generated by coughs.

With the exception of the remarkable lower protection of FFP1 in comparison to FFP2 while coughing, our results suggest that the effectiveness of the masks should mostly be considered based on the generation of secondary jets rather than on the ability to mitigate the front throughflow. For coughing, our findings reveal that all masks and shields enable a reduction of at least 63% of the distance of the filtered front throughflow. However, this estimate is highly conservative because the true maximum jet distance of the unfiltered cough cannot be accurately measured in our images. This is because the air emitted from the subject mixes with the ambient and cools significantly, as the jet travels away from the subject. After approximately 1.2 m, the schlieren signal can no longer be resolved, but the jet up to 1.2 m is still rapidly progressing. The schlieren signal with face covers does not suffer from signal loss in the same manner. Other studies have reported coughing jets extending 2-3 m [21]. If the coughs in this work also extend this distance, our findings would suggest that the masks and shields used in this study reduce the front throughflow distance by approximately 86%.

Our primary metric for comparing the efficacy of face covers is the extent to which they mitigate the spread of exhaled air. However, face coverings not only are highly effective in mitigating the spread of exhaled air, but they also filter the exhaled air. The filtering efficacy is beyond the scope of this study, but we should highlight that air redirected by valves and shields, and leaked through gaps between the face and the mask, is likely to be less filtered than the air flowing through the mask fabric. Redirected air is still partially filtered because largest droplets, which do not follow the air trajectory, are likely to land on the face covering. However, we should assume that the filtering efficacy is lower than flowing through the fabric. Therefore, for the same flow rate and distance travelled, redirected airflows can be more dangerous than filtered throughflow.

The respirator, for example, has a valve system that allows exhaled air to bypass the filter. Hence, air is redirected but not fully filtered. This valve system, which is common to several mask types, is clearly ineffective in preventing virus dispersion when worn by infected people, and it should only be considered to provide protection for healthy wearers against potentially infected people, and only when the specific filter is capable to stop virus particles.

Surgical and handmade masks, and face shields, generate significant leakage jets that have the potential to disperse virus-laden fluid particles by several metres. The different nature of the masks and shields makes the direction of these jets difficult to predict, but the directionality of these jets should be a main design consideration for these covers. They all showed an intense backward airflow for heavy breathing and coughing conditions. It is important to be aware of this flow, to avoid a false sense of security that may arise when standing to the side of, or behind, a person wearing a surgical, or handmade mask, or shield. This is of relevance given the potential for some wearers of surgical masks to turn their face to the side when they cough, during face-to-face interactions with a colleague. In doing so, our data show that there is a risk that this backward jet is directed closer to a person standing in front of the wearer. Additionally, clinicians working around a patient, in the confined space around an intensive care bed or an operating table, are likely to be exposed to these side and backward leakage jets from surgical masks worn by colleagues.

The handmade mask and the face shield opened at the bottom (UoE lightweight 3D-printed face shield), showed an intense downward jet. These results suggest that face shields could be counterproductive in a supine position to reduce outward transmission of pathogens from patients, because virus-laden fluid particles could be redirected over the body of the person and then towards the centre of the room.

Our conclusions confirm the findings of Kähler and Hain [10] that all face coverings are effective in slowing down the front throughflow and, in turn, the horizontal distance travelled by the aerosol. They also confirm the numerical simulations of Dbouk and Drikakis [11], which highlighted the significant leakage jets of non-graded masks. In addition, they complement these studies by providing a detailed description of the different leakage jets for quiet and heavy breathing, and coughing. It is interesting to observe that the first draft of these three complementary works were submitted within four days to each other without knowledge of each other work (a preprint of this paper was submitted to arXiv.org on 19 May, Kähler and Hain [10] was submitted to the Journal of Aerosol Science the following day, and Drikakis [11] was submitted to Physics of Fluids on 4 days later).

Our simulation of an Aerosol Generating Procedure (AGP) revealed that the air jet generated by a cough during the extubation manoeuvre is directed almost vertically in front of the clinicians' faces. This suggests that as long as the patient does not move their head and the clinician does not obstruct or redirect the cough jet with their hand during extubation, the extubating clinician positioned behind the head of the patient is likely to be outside the direct cough jet. While this is achievable during simulations using manikins, it is not always achievable in real-life clinical settings especially when extubating uncooperative patients, infants and young children who cannot follow command, or those with cognitive impairments. Furthermore, extubation in the intensive care unit often involves one clinician standing at the bedside assisting the extubating clinician at the head end behind the patient. The assisting clinician at the bedside may be within the direct cough jet. Patients cannot wear face coverings during AGP, hence it would be desirable to have a reliable and consistent method to remove the direct cough jet during extubation in real-life clinical settings to counteract any potential redirections of the cough jet from patient movements etc. For example, for AGPs, a suitable PPE would be an aerosol extractor to provide a primary layer of protection for clinicians, performing AGPs and tested it on the above model. Work is ongoing to provide this aerosol extractor for clinical use.

A limitation of our experimental setup is that it does not reveal which is the absolute maximum distance that a virus-laden fluid particle can travel, nor how the concentration of these particles varies spatially and temporally. Furthermore, we tested only one person and one manikin while coughing. Hence, these results do not allow to conclusively identify safe distances for different PPEs. However, because droplet evaporation and aerosol buoyancy vary substantially with temperature and humidity, and human breath and cough characteristics also vary significantly, it might not be meaningful searching for conclusive values. Other limitations are that the measured signal is an integral value along line of sight from the camera and thus there is no information on the velocity distribution along this line; the signal is correlated with the temperature gradient and thus its intensity decreases as the jet progresses; the temperature of the masks may affect the temperature of the jet and thus its signal (e.g. the respirator second breaths was more visible than the first one, suggesting that it had absorbed heat from first breath). Within the limits of this work, these results revealed some key relative differences between face covers that can aid policy makers to make informed decisions and PPE developers to improve their product effectiveness by design.

## IV. Materials and Method

Schlieren is an optical technique that provides visualisation of density variations within a flowing medium [27]. These density variations yield refractive index gradients, which refract (i.e. bend) light rays that pass through the medium. A spatial filter is used to segregate the variations of refracted light, providing visualisation of these optical phase disturbances (see also Supplementary Materials, Data Processing).

Schlieren imaging has been used in the past to examine airflows associated with human coughs [13], [16], [28], where the density gradients are produced by temperature differences between a human's breath and the surrounding air. This study utilises a variant of schlieren imaging called background oriented schlieren (BOS), also known as synthetic schlieren [17]. BOS visualises density gradients as refracted light rays distort a patterned background within the image [29]. A reference image without the schlieren object in the field of view provides an image of the stationary patterned background. The apparent local distortion of the patterned background is determined by comparing the schlieren object images to the reference image. This apparent distortion (i.e. the schlieren), are directly related to the magnitude of the density gradient, and yields a 2D image of the density gradients associated with the schlieren object [30], [31]. See also Supplementary Materials, Instrumentation, and Data Processing.

BOS imaging was performed for a human's quiet breathing, heavy breathing and cough (Supplementary Materials, Breathing Patterns). Tests were also conducted using the anatomically realistic adult medical simulation manikin Resusci Anne QCPR with accurate upper airway anatomy/morphology, specifically designed for training in upper airway procedures (Supplementary Materials, Cough Generator).

The experimental setup used for BOS imaging is shown in Fig. 15. The subject of interest (human or manikin) was positioned equidistant ($L/2$) between the camera and patterned background. A new set of reference images were recorded before the start of each experiment.

Figure 16 shows the seven types of face protections that were studied: FFP1, FFP2, respirator, surgical mask, handmade mask, a heavy-duty commercial face shield, and a lightweight face shield based on a 3D printed headband, made at the University of Edinburgh's (UoE) School of Engineering according to an open-source design made available by other groups[1,2,3,4].

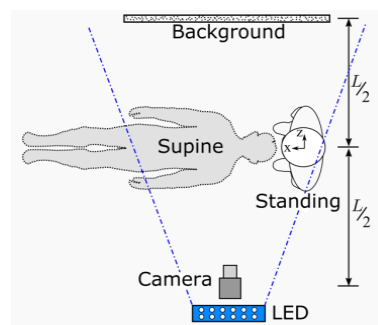

Fig. 15. Experimental setup and reference system for standing and supine configurations.

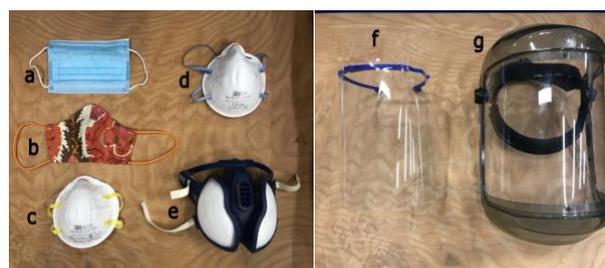

Fig. 16. Different face covers tested: (a) surgical mask; (b) handmade mask; (c) FFP1; (d) FFP2; (e) respirator; (f) university-made lightweight face shield; (g) commercially heavy-duty face shield.

## V. Conclusions

We used a background oriented schlieren technique to investigate the airflow ejected by a person quietly and heavily breathing, and coughing. We tested the effectiveness of different face covers including FFP2 and FFP1 masks, a respirator, a surgical and a handmade mask, and two types of face shields. Finally, we simulated an aerosol generating procedure demonstrating the extent of aerosol dispersion.

For coughing, all face covers, with the exception of the respirator, allow a reduction of the front flow through jet by at least 63% but maybe as high as 86% if the unfiltered cough reaches distances beyond the measurable limits in this work. For the FFP1 and FFP2 masks, which do not have the valve system, the airflow is pushed through the mask material and the front throughflow does not extend by more than $40\pm2$ cm for the FFP1 mask and $11\pm2$ cm for FFP2 mask. However, if these masks are not correctly fitted though, leaking jets are formed. These jets not only can travel significant distances (beyond the boundaries of our field of view), but are also only partially filtered because they do not flew through the mask fabric.

Surgical and handmade masks, and face shields, generate significant backward leakage jets that have the potential to disperse virus-laden fluid particles by several metres. The handmade mask and the face shield opened at the bottom, showed an intense downward jet. The different nature of the masks and shields makes the direction of these jets difficult to be predicted, but the directionality of these jets should be a main design consideration for these covers.

---

[1] https://3dverkstan.se/protective-visor
[2] https://3dprint.nih.gov/discover/3dpx-013306
[3] https://www.edinburghems.com
[4] https://open.ed.ac.uk/3d-visor-models

Finally, visualisation of the air jets during an extubation demonstrates the urgent need to develop technology and procedures to mitigate the risks of infection for the clinicians and other people in the room during and for a period of time after AGPs.

SUPPLEMENTARY MATERIALS

In Supplementary Materials include a detailed description of the methodology, including (I) Instrumentation, (II) Data Processing, (III) Breathing Patterns, and (IV) Cough Generator. Section V includes the three tables with the front flowthrough and the main leakage flow for quiet and heavy breathing, and coughing, respectively. The front flowthrough angle and direction for every face covering were also presented graphically in Fig. 7, 8 and 9. Furthermore, metadata (>250 GB) is available on the Edinburgh DataShare (https://datashare.is.ed.ac.uk/handle/10283/3636). These include, the spirometry tests for the human volunteer and, for each of the 244 tests undertaken, the measured raw data (camera pictures) and the processed data showing the displacements for each frame, and a video for ease of visualisation. In addition, for selected tests, including all of those for which quantitative data is provided in the paper, there are images with annotated measurements.

ACKNOWLEDGMENT

The authors are grateful to LaVision for borrowing of further equipment, which enabled the BOS measurements, to Dr Nandita Chinchankar (Paediatric Intensive Care Unit, Royal Hospital for Sick Children, Edinburgh) for assisting the simulation of the extubation procedure, and the Resuscitation Department at the Royal Infirmary of Edinburgh for providing the simulation manikin.

AUTHORS CONTRIBUTION

Peterson, Pisetta, Pavar, Akhtar, Menoloascina, Mangano and Viola designed the experiments. Mangano and Menolascina developed the cough simulator. Peterson, Pisetta, Pavar, Akhtar undertook the measurements. Pisetta, Pavar and Nila processed the data. McDougall led the simulation of the extubation procedure and conducted the spirometry. Viola coordinated the project and, together with Peterson, Pisetta and Pavar wrote the first draft of the manuscript. All authors advised on the design of experiments, the results interpretation, and edited and approved the manuscript.

REFERENCES

[1] D. K. Chu *et al.*, "Physical distancing, face masks, and eye protection to prevent person-to-person transmission of SARS-CoV-2 and COVID-19: a systematic review and meta-analysis.," *Lancet*, vol. 6736, no. 20, 2020.
[2] A. Rodriguez-Palacios, F. Cominelli, A. R. Basson, T. T. Pizarro, and S. Ilic, "Textile Masks and Surface Covers—A Spray Simulation Method and a 'Universal Droplet Reduction Model' Against Respiratory Pandemics," *Front. Med.*, vol. 7, no. May, pp. 1–11, 2020.
[3] D. Wang *et al.*, "Selection of homemade mask materials for preventing transmission of COVID-19: A laboratory study," *PLoS One*, vol. 15, no. 10 October, pp. 1–13, 2020.
[4] H. Ueki *et al.*, "Effectiveness of Face Masks in Preventing Airborne Transmission of SARS-CoV-2," *mSphere*, vol. 5, no. 5, pp. 1–5, 2020.
[5] S. Asadi, C. D. Cappa, S. Barreda, A. S. Wexler, N. M. Bouvier, and W. D. Ristenpart, "Efficacy of masks and face coverings in controlling outward aerosol particle emission from expiratory activities.," *Sci. Rep.*, vol. 10, no. 1, p. 15665, 2020.
[6] L. Bandiera *et al.*, "Face Coverings and Respiratory Tract Droplet Dispersion," *Prepr. to Appear Medarxiv*, pp. 1–12, 2020.
[7] S. Verma, M. Dhanak, and J. Frankenfield, "Visualizing the effectiveness of face masks in obstructing respiratory jets," *Phys. Fluids*, vol. 061708, no. 32, pp. 1–7, 2020.
[8] S. Verma, M. Dhanak, and J. Frankenfield, "Visualizing droplet dispersal for face shields and masks with exhalation valves," *Phys. Fluids*, vol. 091701, no. 32, 2020.
[9] C. J. Kähler and R. Hain, "Flow analyses to validate SARS-CoV-2 protective masks," 2020.
[10] C. J. Kähler and R. Hain, "Fundamental protective mechanisms of face masks against droplet infections," *J. Aerosol Sci.*, vol. 148, no. May, 2020.
[11] T. Dbouk and D. Drikakis, "On respiratory droplets and face masks," *Phys. Fluids*, vol. 32, no. 6, 2020.
[12] G. S. Settles, "Fluid Mechanics and Homeland Security," *Annu. Rev. Fluid Mech.*, vol. 38, no. 1, pp. 87–110, Jan. 2006.
[13] J. W. Tang and G. S. Settles, "Coughing and Aerosols," *N. Engl. J. Med.*, vol. 359, no. 15, p. e19, Oct. 2008.
[14] J. W. Tang *et al.*, "Observing and quantifying airflows in the infection control of aerosol- and airborne-transmitted diseases: an overview of approaches," *J. Hosp. Infect.*, vol. 77, no. 3, pp. 213–222, Mar. 2011.
[15] J. W. Tang *et al.*, "Airflow Dynamics of Human Jets: Sneezing and Breathing - Potential Sources of Infectious Aerosols," *PLoS One*, vol. 8, no. 4, p. e59970, Apr. 2013.
[16] J. W. Tang, T. J. Liebner, B. A. Craven, and G. S. Settles, "A schlieren optical study of the human cough with and without wearing masks for aerosol infection control," *J. R. Soc. Interface*, vol. 6, no. SUPPL. 6, pp. 727–736, 2009.
[17] S. B. Dalziel, G. O. Hughes, and B. R. Sutherland, "Whole-field density measurements by 'synthetic schlieren,'" *Exp. Fluids*, vol. 28, no. 4, pp. 322–335, Apr. 2000.
[18] F. A. Berlanga *et al.*, "Influence of the geometry of the airways on the characterization of exhalation flows. Comparison between two different airway complexity levels performing two different breathing functions," *Sustain. Cities Soc.*, vol. 53, no. September 2019, 2020.
[19] G. Seminara, B. Carli, G. Forni, S. Fuzzi, A. Mazzino, and A. Rinaldo, "Biological fluid dynamics of airborne COVID-19 infection," *Rend. Lincei*, vol. 31, no. 3, pp. 505–537, 2020.
[20] L. Bourouiba, "Turbulent Gas Clouds and Respiratory Pathogen Emissions," *JAMA*, pp. 1–2, Mar. 2020.
[21] L. Bourouiba, E. Dehandschoewercker, and J. W. M. Bush, "Violent expiratory events: On coughing and sneezing," *J. Fluid Mech.*, vol. 745, pp. 537–563, Apr. 2014.
[22] M. Vansciver, S. Miller, and J. Hertzberg, "Particle Image Velocimetry of Human Cough," *Aerosol Sci. Technol.*, vol. 45, no. 3, pp. 415–422, Feb. 2011.
[23] S. Zhu, S. Kato, and J.-H. Yang, "Study on transport

characteristics of saliva droplets produced by coughing in a calm indoor environment," *Build. Environ.*, vol. 41, no. 12, pp. 1691–1702, Dec. 2006.

[24] J. K. Gupta, C.-H. Lin, and Q. Chen, "Characterizing exhaled airflow from breathing and talking," *Indoor Air*, vol. 20, no. 1, pp. 31–39, Feb. 2010.

[25] M. Varshney, R. Kumar, K. Sharma, and P. Varshney, "Appropriate depth of placement of oral endotracheal tube and its possible determinants in Indian adult patients," *Indian J. Anaesth.*, vol. 55, no. 5, p. 488, 2011.

[26] J. W. Cavo, "True Vocal Cord Paralysis Following Intubation," *Laryngoscope*, vol. 95, no. 11, p. 1352???1359, Nov. 1985.

[27] G. S. Settles, *Schlieren and Shadowgraph Techniques: Visualizing Phenomena in Transparent Media*. Berlin: Springer, 2001.

[28] T. A. Khan, H. Higuchi, D. R. Marr, and M. N. Glauser, "Unsteady flow measurements of human micro environment using time-resolved particle image velocimetry," *Proc. Room Vent 2004*, 2004.

[29] H. Richard and M. Raffel, "Principle and applications of the background oriented schlieren (BOS) method," *Meas. Sci. Technol.*, vol. 12, no. 9, pp. 1576–1585, Sep. 2001.

[30] M. J. Hargather and G. S. Settles, "Natural-background-oriented schlieren imaging," *Exp. Fluids*, vol. 48, no. 1, pp. 59–68, Jan. 2010.

[31] S. Tokgoz, R. Geisler, L. J. A. van Bokhoven, and B. Wieneke, "Temperature and velocity measurements in a fluid layer using background-oriented schlieren and PIV methods," *Meas. Sci. Technol.*, vol. 23, no. 11, p. 115302, Nov. 2012.

# Supplementary Materials

Face Coverings, Aerosol Dispersion and Mitigation of Virus Transmission Risk

I. M. Viola*, B. Peterson, G. Pisetta, G. Pavar, H. Akhtar, F. Menoloascina, E. Mangano, K. E. Dunn, R. Gabl, A. Nila , E. Molinari, C. Cummins, G. Thompson, T. Y. M. Lo, F. C. Denison, P. Digard, O. Malik, M. J. G. Dunn, C. M. McDougall, F. V. Mehendale

## I. INSTRUMENTATION

A high-speed CMOS camera (VEO 710L, Vision Research) with 1280 x 800 pixels$^2$ was used for recording. The camera operated in single-frame mode with a repetition rate of 100 Hz and exposure time of 9996.6 $\mu$s. Hence, each image is identified by a frame number. The camera was equipped with a 50 mm Nikon lens operating with f# = 11.

The camera imaged onto a patterned background placed at a distance $L$ = 2200 mm from the camera. The patterned background was printed onto a matte poster-board (1400 x 1000 mm$^2$). The background was generated using LaVision's Random Pattern Generator 1.3 software with dot size of 2.2 mm and a minimum dot distance of 1.1 mm. A blue LED-Flashlight 300 (LaVision) was used to illuminate the background. This LED unit is comprised of 72 high-power LEDs and operated with a pulse duration of 300 $\mu$s. A piece of sanded-down transparent acetate was placed in front of the LED unit to diffuse the illuminated light. A programmable timing unit (LaVision) was used to synchronize the LED light source with the high-speed camera. Acquisition and analysis of the BOS images were performed using commercial software (DaVis 10.1 by LaVision).

## II. DATA PROCESSING

The raw images were processed using the LaVision's DaVis 10 software. First, a time filter was applied to the raw images in a pre-processing stage to mitigate the effects of shot noise. The filter computes the intensity for each pixel from five consecutive images and assigns the maximum value to the pixel in the filtered image. The flow field was then visualised by evaluating the BOS displacement, which follows the motion of the dot pattern in the background on each image with respect to the reference image recorded before the test, where no flow was present (natural airflow in the room was minimised as far as possible). The software computes an average displacement value of the dots within a square window, called a subset, chosen in this case to be of 19 x 19 pixels$^2$. After calculating the displacement in the current subset, the process was repeated by shifting the subset by five pixels, until the entire image was scanned. The image displacement data was fitted with a first order polynomial which is then subtracted from the original BOS displacement map. This allows to eliminate the global bias in the BOS field which might be occurring due to motion and vibration or due to overall illumination variations. The contours of the resulting computed displacement are used to visualise the air flow.

## III. BREATHING PATTERNS

Spirometry (Nuvoair Next) was performed on the person who performed all the tests shown in this paper, in line with ATS/ERS standards for spirometry [1]. The person is a co-author of this paper and he gave informed consent to undertake these measurements. These measurements were conducted by an Intensive Care Specialist using the Nuvoair Air Next, a CE certified and FDA cleared portable spirometer. Detailed results are available as Supplementary Material (Spirometry Results). Measurements of forced vital capacity were 1.29, 4.64, 1.65 litres over durations of 3.25, 2.09, 1.32 s for quite breathing, heavy breathing and coughing, respectively. The forced expired volumes in one second were 0.90, 4.33, 1.63 litres, respectively. The peak expiratory flows were 1.21, 8.99, 5.86 litres/second, respectively.

We also perform a reduced-order characterisation of the breathing patterns using BOS. At the beginning of a cough event, as air exits the subject's mouth, the schlieren signal is the strongest. Progression of the leading edge of the gas cloud can be clearly tracked within the first 1-3 frames when the schlieren signal is first observed near the subject's mouth. An estimate of the air velocity is calculated by recording the distance between the leading edge of the signal within successive frames and dividing by the time elapsed between the frames. This velocity estimation is inherently averaged across the line of sight of the camera. The velocities measured near the subject's mouth approximate the average peak velocity at the beginning of a cough. Conversely, for quiet and heavy breathing, we estimated the velocity as the maximum distance covered by the leading edge of the gas cloud within the field of view over the time it took to reach that point. This reduces the global bias in the BOS field, which may occur due to motion and vibration or due to overall illumination variations.

## IV. Cough Generator

The manikin enables repeatability of the tests when comparing different face coverings and it allowed increasing the temperature of the airflow to measure the displacement of the jet further downstream by moving the field of view.

The manikin's trachea is connected to a built-in-house breath/cough simulator. The system is designed to replicate the key aspects of human breath and cough by generating an air flow which is then released to the manikin's trachea at pre-defined time intervals. Air flow is generated from an air compressor for low velocity flow, i.e. simulation of quite breathing; while compressed air from a cylinder is used to simulate cough events. The dual air source allows to set the outlet air velocity in the whole range of velocities reported for human breath and cough. The system is connected to the mannikin's trachea via a remotely-controlled solenoid valve. In the case of breath simulation the valve is kept open for the entire duration of the experiment, while to simulate cough events the is controlled using a timed sequence which allows to open the valve for very short time intervals. This allows to simulate cough events 200 ms, in line with the literature [2].

The air entering the manikin's trachea is pre-heated using electrical heating elements to ensure an outlet flow from the manikin mouth of about 40°C. The air temperature within the room was 20-22°C. The plastic tube connecting the manikin to the breath simulator is insulated, to minimise the generation of noise during BOS measurements.

The manikin's mouth has fixed shape and an aperture of about 6.4 cm$^2$, while the real person mouth was open to about 3 cm$^2$ (note that the manikin and the person were used to perform different sets of tests and thus this difference is irrelevant). The manikin's nasal airway is sealed to prevent leakage.

## V. Front Throughflow and Main Leakage for Different Face Covers

**Table SM-I.** Quiet breath, human, sideways (95% CI < $\pm 5°$, $\pm 2$ cm).

| Mask | Test | Front throughflow | | | | Brow jet | | | |
|---|---|---|---|---|---|---|---|---|---|
| | | Frame | Angle (deg) | Distance (cm) | H Dist. (cm) | Frame | Angle (deg) | Distance (cm) | V Dist (cm) |
| none | 204 | 557 | 1 $\pm 5$ | >56 | >56 | not discernible | | | |
| surgical | 98 | 87 | 48 $\pm 5$ | 9 $\pm 2$ | 6 $\pm 2$ | 232 | 88 $\pm 5$ | >21 $\pm 2$ | >21 $\pm 2$ |
| FFP1 | 101 | 182 | 46 $\pm 5$ | 8 $\pm 2$ | 5 $\pm 2$ | not discernible | | | |
| FFP2 | 103 | 104 | -15 $\pm 5$ | 3 $\pm 2$ | 3 $\pm 2$ | not discernible | | | |
| handmade | 105 | not discernible | | | | not discernible | | | |
| respirator | 107 | 57 | -48 $\pm 5$ | 15 $\pm 2$ | 10 $\pm 2$ | not discernible | | | |
| heavy-duty commercial face shield | 109 | not discernible | | | | not discernible | | | |
| UoE lightweight 3DP face shield + opaque cover | 133 | not discernible | | | | not discernible | | | |

**Table SM-2.** Heavy breath, human, sideways (95% CI < $\pm 5°$, $\pm 2$ cm).

| Mask | Test | Front throughflow | | | | Brow jet | | | |
|---|---|---|---|---|---|---|---|---|---|
| | | Frame | Angle (deg) | Distance (cm) | H Dist. (cm) | Frame | Angle (deg) | Distance (cm) | V Dist (cm) |
| none | 198 | 181 | -7 $\pm 5$ | >55 | >55 | not discernible | | | |
| surgical | 99 | 80 | -24 $\pm 5$ | 28 $\pm 2$ | 25 $\pm 2$ | 67 | 58 $\pm 5$ | 18 $\pm 2$ | 15 $\pm 2$ |
| FFP1 | 102 | 197 | -18 $\pm 5$ | 20 $\pm 2$ | 19 $\pm 2$ | not discernible | | | |
| FFP2 | 104 | 95 | 12 $\pm 5$ | 20 $\pm 2$ | 19 $\pm 2$ | not discernible | | | |
| handmade | 106 | 41 | -11 $\pm 5$ | 14 $\pm 2$ | 14 $\pm 2$ | 32 | 32 $\pm 5$ | 20 $\pm 2$ | 11 $\pm 2$ |
| respirator | 36 | 135 | -54 $\pm 5$ | >27 | >16 | not discernible | | | |
| heavy-duty commercial face shield | 110 | 206 | -24 $\pm 5$ | 14 $\pm 2$ | 12 $\pm 2$ | not discernible | | | |
| UoE lightweight 3DP face shield + opaque cover | 134 | not discernible | | | | not discernible | | | |

**Table SM-III.** Coughing, manikin, sideways (95% CI < $\pm 5°$, $\pm 2$ cm).

| Mask | Test | Front throughflow | | | | Crown jet | | | |
|---|---|---|---|---|---|---|---|---|---|
| | | Frame | Angle (deg) | Distance (cm) | H Dist. (cm) | Frame | Angle (deg) | Distance (cm) | V Dist (cm) |
| none | 253, 254, 255 | 84, 72, 172 | -10±5 | 114±2 | 112±2 | not discernible | | | |
| surgical | 256 | 77 | -24±5 | 36±2 | 33±2 | 17 | 114±5 | >17 | >15 |
| FFP1 | 257 | 236 | 16±5 | 42±2 | 40±2 | 64 | 118±5 | >18 | >15 |
| FFP2 | 258 | 119 | -22±5 | 12±2 | 11±2 | 74 | 121±5 | >19 | >16 |
| handmade | 259 | 123 | 14±5 | 23±2 | 23±2 | 61 | 123±5 | >19 | >16 |
| respirator | 260 | 59 | -61±5 | >31 | >15 | 57 | 99±5 | >11 | >11 |
| heavy-duty commercial face shield | 261 | not discernible | | | | 70 | 82±5 | 13±2 | >6 |
| UoE lightweight 3DP face shield + opaque cover | 262 | not discernible | | | | 80 | 38±5 | 9±2 | >4 |

## Metadata

Metadata (>250 GB) is available on the Edinburgh DataShare at https://datashare.is.ed.ac.uk/handle/10283/3636. Data include, the spirometry tests for the human volunteer and, for each of the 244 tests undertaken, the measured raw data (camera pictures) and the processed data showing the displacements for each frame, and a video for ease of visualisation. In addition, for selected tests, including all of those for which quantitative data is provided in the paper, there are images with annotated measurements.

## References

[1] M. R. Miller, "Standardisation of spirometry," *Eur. Respir. J.*, vol. 26, no. 2, pp. 319–338, Aug. 2005.

[2] J. K. Gupta, C.-H. Lin, and Q. Chen, "Flow dynamics and characterization of a cough," *Indoor Air*, vol. 19, no. 6, pp. 517–525, Dec. 2009.